\documentclass[showpacs,preprintnumbers,amsmath,amssymb,prb]{revtex4}

\usepackage{graphicx}
\usepackage{dcolumn}
\usepackage{bm}

\begin{document}

\title{Novel elementary excitations in spin-$\frac12$ antiferromagnets on the triangular lattice}

\author{A.\ V.\ Syromyatnikov}
\email{asyromyatnikov@yandex.ru}
\affiliation{National Research Center "Kurchatov Institute" B.P.\ Konstantinov Petersburg Nuclear Physics Institute, Gatchina 188300, Russia}
\affiliation{National Research University for Information Technology, Mechanics and Optics (ITMO), St.\ Petersburg, 197101, Russia}

\date{\today}

\begin{abstract}

We discuss spin-$\frac12$ Heisenberg antiferromagnet on the triangular lattice using the recently proposed bond-operator technique (BOT). We use the variant of the BOT which takes into account all spin degrees of freedom in the magnetic unit cell containing three spins. Apart from conventional magnons known from the spin-wave theory (SWT), there are novel high-energy collective excitations in the BOT which are built from high-energy excitations of the magnetic unit cell. We obtain also another novel high-energy quasiparticle which has no counterpart not only in the SWT but also in the harmonic approximation of the BOT. All observed elementary excitations produce visible anomalies in dynamical spin correlators. We show that quantum fluctuations considerably change properties of conventional magnons predicted by the SWT. The effect of a small easy-plane anisotropy is discussed. The anomalous spin dynamics with multiple peaks in the dynamical structure factor is explained that was observed recently experimentally in $\rm Ba_3CoSb_2O_9$ and which the SWT could not describe even qualitatively.

\end{abstract}


\maketitle

\section{Introduction}

Plenty of collective phenomena are discussed in the modern theory of many-body systems in terms of appropriate elementary excitations (quasiparticles). \cite{agd,stphys,auer,zinn,sachdev} According to the quasiparticle concept, each weakly excited state of a system can be represented as a set of weakly interacting quasiparticles carrying quanta of momentum and energy. Thus, the search and characterization of elementary excitations is of fundamental importance. Because poles of Green's functions are determined by spectra of elementary excitations, quasiparticles produce peaks in dynamical correlators which can be observed experimentally and numerically. However some peaks can be smeared due to their small spectral weights (small residues of the corresponding poles), insufficient experimental resolution, finite-size effects (in numerical studies), and/or a finite quasiparticle damping. Besides, some anomalies in observable quantities can be not of single-quasiparticle nature originating from continuums of excitations. Then, the interpretation of numerical and experimental data relies heavily on conclusions of existing analytical approaches operating with suitable elementary excitations. 

It can be stated that properties are well understood of long-wavelength elementary excitations (magnons) in ordered phases of quantum spin systems. \cite{auer,chak,andreev,nlsmtriang,hydro,hydro2} However, there is a growing number of evidences that in (quasi-)two-dimensional collinear and non-collinear quantum systems standard analytical methods do not describe properly short-wavelength spin excitations. 

For example, a mysterious anomaly of the magnon spectrum near the momentum ${\bf k}=(\pi,0)$ was found experimentally \cite{chris1,piazza} and numerically \cite{qmc2,sand,ser,spinon,piazza,pow1,pow2} in spin-$\frac12$ Heisenberg antiferromagnet (HAF) on the square lattice. Besides, a distinct continuum of excitations arises in the transverse dynamical structure factor (DSF) at ${\bf k}=(\pi,0)$ which has the form of a high-energy tail of the one-magnon peak (similar features were observed also in layered cuprates \cite{cupr1,cupr2,cupr3}). This continuum was ascribed to a magnon instability at ${\bf k}=(\pi,0)$ with respect to a decay either into two spinons \cite{spinon,piazza,spinon2,spinon3} or into a Higgs excitation and another magnon \cite{pow1,pow2,moess}. 

An even more exotic picture was discovered numerically in spin-$\frac12$ HAF on the square lattice in strong magnetic field: not far from the saturation field, a large number of peaks (instead of one magnon peak) appear in dynamical spin correlators at a given momentum. \cite{olav,magfail2} These anomalies were interpreted as an indication of a short-wavelength magnons instability observed self-consistently in the spin-wave theory (SWT) in the first order in $1/S$, where $S$ is the spin value. \cite{olav,zhito}

\begin{figure}
\includegraphics[scale=1.0]{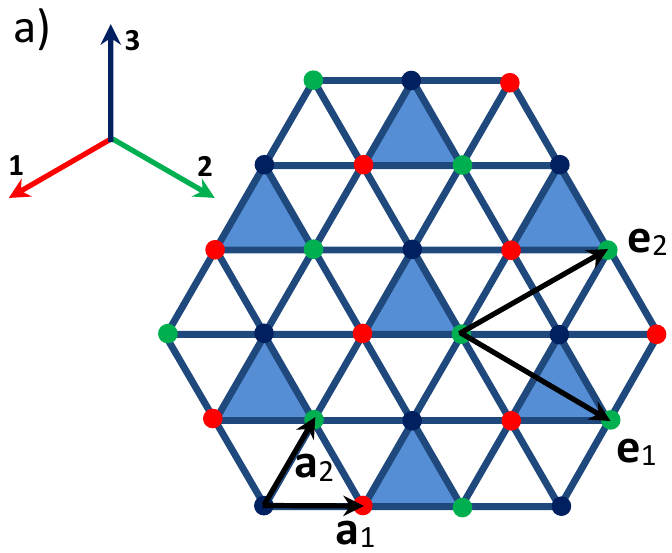}
\includegraphics[scale=0.9]{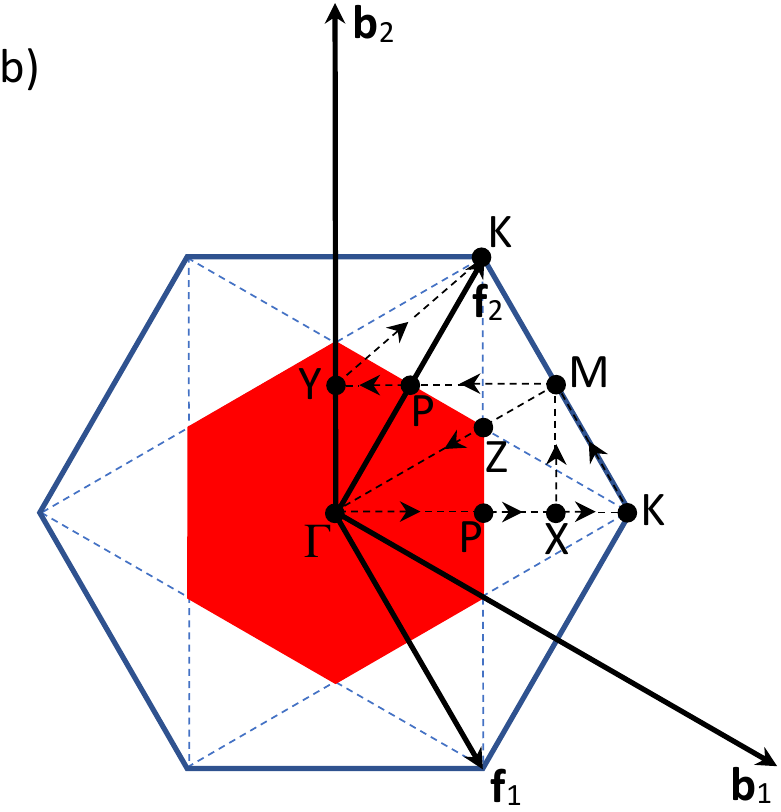}
\caption{
(a) Antiferromagnet on the triangular lattice with three spins in the magnetic unit cell. Sites are distinguished by color belonging to three magnetic sublattices in the magnetically ordered ground state. Translation vectors are shown of the crystal (${\bf a}_{1,2}$) and of the magnetic lattice (${\bf e}_{1,2}$).
(b) Crystal (blue hexagon) and magnetic (red hexagon) Brillouin zones. Translation vectors are depicted of the crystal (${\bf b}_{1,2}$) and of the magnetic (${\bf f}_{1,2}$) reciprocal lattices.
\label{lattfig}}
\end{figure}

Then, in a series of recent inelastic neutron scattering experiments carried out in $\rm Ba_3CoSb_2O_9$, the complete inability was demonstrated of standard theoretical approaches to describe short-wavelength spin excitations in spin-$\frac12$ HAFs on the triangular lattice. \cite{triang1,bacoprl,bacoprb} In particular, at least four peaks can be distinguished in experimentally found DSFs at $M$ point (see Fig.~\ref{lattfig}) of the Brillouin zone (BZ), whereas the SWT predicts only two magnon peaks and a high-energy continuum of excitations. \cite{chub_triang,zh_triang,zhito} 
Recent application of the Shwinger boson approach to this problem reproduces qualitatively high-energy peculiarities in experimental data. \cite{batista} A quantitative agreement with the experiment is achieved recently in a numerical consideration using the tensor network renormalization group method. \cite{navy} The resonating valence bond (RVB) physics was invoked recently in the description of the ground state and the spin dynamics of this system. \cite{spinontri} Despite certain success in the description of the low-energy spin excitations around $M$ point, the RVB theory failed to discern the experimentally observed anomalies in the high-energy spectral continuum. The anomalous dynamics in $\rm Ba_3CoSb_2O_9$ was explained in Ref.~\cite{bacoprb} only phenomenologically.

Notice also that in view of recent findings that short-wavelength spin excitations can play an important role in the spin-fluctuation-mediated pairing mechanism in high-temperature superconductors, \cite{Tacon} the clarification of the nature of short-wavelength magnons in quantum low-dimensional spin systems would be of broad importance.

We have proposed recently and tested on a number of spin systems a new method based on the bond representation of spin-$\frac12$ operators in terms of Bose operators. \cite{ibot,aktersky,iboth} This bond-operator technique (BOT) is suitable for describing both magnetically ordered and disordered phases (and transitions between them). The idea of the BOT is to increase the unit cell and to construct a representation of all spins in it via Bose operators which create or annihilate quantum states of the whole unit cell. It is clear that along with common quasiparticles (magnons or triplons), there are extra bosons in the spin representation which describe elementary excitations arising in conventional approaches as bound states of magnons or triplons. We have developed a general procedure in Ref.~\cite{ibot} for constructing the bosonic spin representations for arbitrary number of spins in the unit cell. There is a formal parameter $n$ in the BOT, the maximum number of bosons which can occupy a unit cell, that allows a regular expansion of physical observables in powers of $1/n$ (physical results correspond to $n=1$). Importantly, the spin commutation algebra is fulfilled for any $n>0$ that guarantees existence of Goldstone excitations in phases with spontaneously broken continuous symmetry in any order in $1/n$. Then, the BOT is very close in spirit to the standard SWT based on the expansion in powers of $1/S$, but it more accurately takes into account short-range spin correlations and makes it possible, along with magnons, to quite simply study high-energy excitations arising in the SWT as bound states of several magnons. In particular, in the BOT with four spins in the unit cell which was suggested for the ordered phase in spin-$\frac12$ HAF, there are separate bosons describing the amplitude (Higgs) excitation and a spin-0 quasiparticle named singlon. \cite{ibot} The latter is responsible for the anomaly in Raman intensity in the $B_{1g}$ symmetry observed, e.g., in layered cuprates. \cite{ibot} By comparison with other available numerical and experimental results obtained in a number of two-dimensional spin models, we demonstrated \cite{ibot,aktersky} that in most cases first $1/n$ corrections make the main renormalization of the staggered magnetization, the ground-state energy, and energies of quasiparticles (similar to the SWT in which first $1/S$ corrections provide in many cases the main renormalization of observable quantities even in two-dimensional systems with $S=1/2$ \cite{monous,auer}). 

Using the BOT with four spins in the unit cell, we quantitatively reproduced in Ref.~\cite{ibot} the anomaly of the magnon spectrum near ${\bf k}=(\pi,0)$ in spin-$\frac12$ HAF on the square lattice and excluded the Higgs-magnon mechanism of the formation of this anomaly. In Ref.~\cite{iboth}, we used the four-spin variant of the BOT to describe numerous anomalies in dynamical spin correlators in spin-$\frac12$ HAF on the square lattice in strong field. A very rare phenomenon was discovered: quantum fluctuations are so strong in this system that these anomalies correspond to poles of Green's functions, which have no counterparts in the semiclassical SWT (i.e., we showed that taking into account self-energy parts in the first order in $1/n$ leads to the appearance of the novel poles). That is, the system contains numerous short-wavelength magnetic excitations (magnons) which have nothing to do with magnons in the SWT. 

In the present paper, we use a three-spin variant of the BOT considered in some detail in Sec.~\ref{secthe BOT} for discussion of spin dynamics in spin-$\frac12$ HAF with a small easy-plane anisotropy on the triangular lattice described by the Hamiltonian
\begin{equation}
\label{ham}
{\cal H} = \sum_{\langle i,j \rangle}	J\left( {\bf S}_i{\bf S}_j - A S_i^y S_j^y \right),
\end{equation} 
where $\langle i,j \rangle$ denote nearest-neighbor sites, the exchange coupling constant $J$ is set to be equal to unity below, and the anisotropy value $A\ge0$. Our theory takes into account all excited states in the magnetic unit cell containing three spins and it respects the symmetry of the magnetic ordering (see Fig.~\ref{lattfig}(a)). We consider static properties in Sec.~\ref{static} of model \eqref{ham} at $A=0$ and show that the $120^\circ$ magnetic ordering is reproduced in the BOT both in the harmonic approximation and in the first order in $1/n$. The ground-state energy and the value of the sublattices magnetization found in the first order in $1/n$ are in good quantitative agreement at $n=1$ with previous analytical and numerical findings.

We calculate the magnon spectrum in Sec.~\ref{dyn} at $A=0$ in the first order in $1/n$. There are seven branches of excitations in the BOT three of which are Goldstone quasiparticles ("low-energy magnons") known from the SWT and the rest four "optical" branches ("high-energy magnons") stem from high-energy excitations of the unit cell. We obtain also the eighth quasiparticle which has no counterparts neither in the SWT no in the harmonic approximation of the BOT. Similar to new elementary excitations obtained in Ref.~\cite{iboth} in the HAF on the square lattice in strong magnetic field (see above), the origin of the eighth quasiparticle is in strong quantum fluctuations in the system. We demonstrate that all observed quasiparticles produce visible anomalies in dynamical spin correlators. Spectra of low-energy magnons are in good agreement with previous numerical results obtained using the series expansion \cite{series_tri} and the dynamical variational Monte Carlo approach \cite{triang3,triang4}. In particular, the BOT reproduces the "roton" minima in the spectrum of the well-defined low-energy magnon around $M$ and $P$ points. We show that quantum fluctuations considerably change properties of conventional magnons predicted by the SWT. In particular, we demonstrate that quantum fluctuations lift a degeneracy of two low-energy magnon branches predicted by the SWT along $\Gamma M$ lines and along blue dashed lines in Fig.~\ref{lattfig}(b) that in turn leads to larger number of peaks in dynamical spin correlators.

The latter conclusion is in quantitative agreement with results of recent experiments in $\rm Ba_3CoSb_2O_9$, as we show in Sec.~\ref{anisotropy} which is devoted to discussion of the small easy-plane anisotropy $A$ in model \eqref{ham} and to comparison of our theory with experiment. In agreement with the conclusion made in the spin-wave analysis \cite{zhito,zh_triang}, we find that even small easy-plane anisotropy reduces considerably the phase space for magnon decay into two other magnons so that four low-energy elementary excitations obtained in the BOT have negligible damping at $A=0.15$. We propose that four anomalies obtained in $\rm Ba_3CoSb_2O_9$ at $M$ point in Ref.~\cite{bacoprl} in the interval 0--3~meV stem from three low-energy magnons and the eighth (novel) quasiparticle. High-energy magnons found in the BOT contribute to a broad anomaly around 3.5~meV observed in $\rm Ba_3CoSb_2O_9$ in Refs.~\cite{triang1,bacoprb}.

Sec.~\ref{conc} contains our conclusion. Three appendixes are added with details of our analysis.

\section{Bond-operator formalism for spin-$\frac12$ magnets on the triangular lattice}
\label{secthe BOT}

\begin{figure}
\includegraphics[scale=0.4]{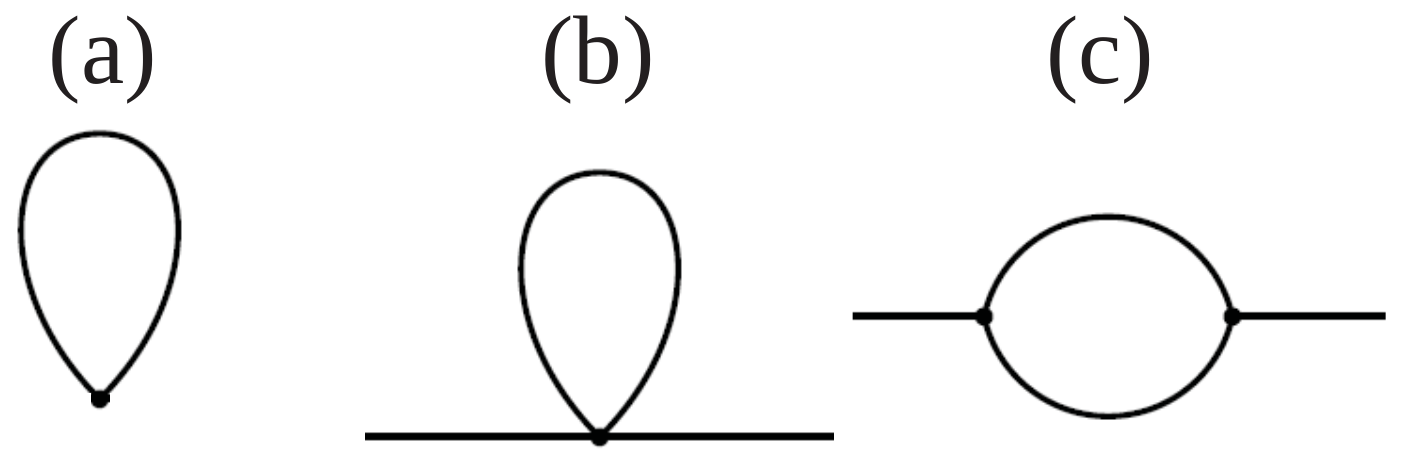}
\caption{Diagrams giving corrections of the first-order in $1/n$ to (a) the ground state energy and the staggered magnetization, and (b), (c) to self-energy parts.
\label{diag}}
\end{figure}

Let us take into account all spin degrees of freedom in the magnetic unit cell containing three spins 1/2 which form a triangle (see Fig.~\ref{lattfig}(a)). The three-spin variant of the BOT can be built according to the general scheme described in detail in Ref.~\cite{ibot}. First, we introduce seven Bose operators in each unit cell which act on eight basis functions of three spins $|0\rangle$ and $|e_i\rangle$ ($i=1,...,7$) according to the rule
\begin{equation}
\label{bosons}
	a_i^\dagger |0\rangle = |e_i\rangle, \quad i=1,...,7,
\end{equation}
where $|0\rangle$ is a selected state playing the role of the vacuum. Suitable basis functions are presented in Appendix~\ref{tecthe BOT}. Then, we build the bosonic representation of spins in the unit cell as it is described in Ref.~\cite{ibot} which turns out to be quite bulky so that we do not present it here. The code in the Mathematica software which generates this representation is presented in the Supplemental Material. There is a formal artificial parameter $n$ in this representation that appears in operator $\sqrt{n-\sum_{i=1}^7a_i^\dagger a_i}$ by which linear in Bose operators terms are multiplied (cf.\ the term $\sqrt{2S-a_i^\dagger a_i}$ in the Holstein-Primakoff representation). It prevents mixing of states containing more than $n$ bosons and states with no more than $n$ bosons (then, the physical results of the BOT correspond to $n=1$). Besides, all constant terms in our representation of spin components are proportional to $n$ whereas bilinear in Bose operators terms do not depend on $n$ and have the form $a_i^\dagger a_j$. We introduce also separate representations via operators \eqref{bosons} for terms ${\bf S}_i{\bf S}_j$ in the Hamiltonian in which $i$ and $j$ belong to the same unit cell. Constant terms in these representations are proportional to $n^2$ and terms of the form $a_i^\dagger a_j$ are proportional to $n$. \cite{ibot} Thus, we obtain a close analog of the conventional Holstein-Primakoff spin transformation which reproduces the commutation algebra of all spin operators in the unit cell for all $n>0$ and in which $n$ is the counterpart of the spin value $S$. In analogy with the SWT, expressions for observables are found in the BOT using the conventional diagrammatic technique as series in $1/n$. This is because terms in the Bose-analog of the spin Hamiltonian containing products of $i$ Bose operators are proportional to $n^{2-i/2}$ (in the SWT, such terms are proportional to $S^{2-i/2}$). For instance, to find the ground-state energy, the staggered magnetization, and self-energy parts in the first order in $1/n$ one has to calculate diagrams shown in Fig.~\ref{diag} (as in the SWT in the first order in $1/S$). 

Our previous applications of the BOT to two-dimensional models well studied before by other numerical and analytical methods show that first $1/n$ terms in most cases give the main corrections to renormalization of observables if the system is not very close to a quantum critical point (similarly, first $1/S$ corrections in the SWT frequently make the main quantum renormalization of observable quantities even at $S=1/2$, Ref.~\cite{monous}). \cite{ibot,aktersky} Importantly, because the spin commutation algebra is reproduced in our method at any $n>0$, the proper number of Goldstone excitations arises in ordered phases in any order in $1/n$ (unlike the vast majority of other versions of the BOT proposed so far \cite{ibot}). Although the BOT is technically very similar to SWT, the main disadvantage of this technique is that it is very bulky (e.g., the part of the Hamiltonian bilinear in Bose-operators contains more than 100 terms) and it requires time-consuming numerical calculation of diagrams. That is why there is a limited number of points on some plots below found in the first order in $1/n$.

It should be noted also that we discussed so far ordered states only in systems on bipartite lattices using the BOT. \cite{ibot,aktersky,iboth} However, this approach can be applied without modifications to magnets with commensurate non-collinear magnetic orderings if the unit cell considered in the BOT is the magnetic unit cell (as it is the case in the present study of the triangular-lattice HAF with the three-spin unit cell). Magnetic sublattices appear inside the unit cell automatically after minimization of the ground state energy with respect to parameters $\alpha$ introduced in Appendix~\ref{tecthe BOT} (see Ref.~\cite{ibot} for extra details).

\section{Static properties. $A=0$.}
\label{static}

As it is shown in Appendix~\ref{tecthe BOT}, there are three parameters $\alpha$ in the three-spin variant of the BOT controlling the mixing of the basis functions. These parameters are series in powers of $1/n$ which should be found by minimization of the ground-state energy (at these values of parameters, linear in Bose operators terms are also vanish in the Hamiltonian, see Ref.~\cite{ibot} for extra discussion). Calculating also the diagram shown in Fig.~\ref{diag}(a), we obtain for the staggered magnetization $\langle S\rangle$ and for the ground state energy per spin ${\cal E}$
\begin{equation}
\label{statics}
	\begin{aligned}
		\langle S\rangle &= 0.465n-0.220 \mapsto 0.245 \mbox{ (at }n=1), \\ 
		{\cal E} &= -0.4247n^2-0.1114n \mapsto -0.5361 \mbox{ (at }n=1)
	\end{aligned}
\end{equation}
which are very close to previous numerical and analytical findings (see, e.g., Ref.~\cite{zh_triang} and Table~III in Ref.~\cite{series_tri}). Notice that mean spin components do reproduce the $120^\circ$ magnetic order in the zeroth and in the first order in $1/n$.

\section{Dynamical properties. $A=0$.}
\label{dyn}

We calculate in this section the dynamical spin susceptibility
\begin{equation}
\label{chi}
\chi({\bf k},\omega) = 
i\int_0^\infty dt 
e^{i\omega t}	
\left\langle \left[ {\bf S}_{\bf k}(t), {\bf S}_{-\bf k}(0) \right] \right\rangle
\end{equation}
and the dynamical structure factor (DSF)
\begin{equation}
\label{dsf}
{\cal S}({\bf k},\omega) = 
\frac1\pi {\rm Im}
\chi({\bf k},\omega),
\end{equation}
where
\begin{equation}
\label{sk}
	{\bf S}_{\bf k} = \frac{1}{\sqrt3} 
	\left( 
	{\bf S}_{1\bf k} + {\bf S}_{2\bf k}e^{-i(k_1+k_2)/3} + {\bf S}_{3\bf k}e^{-i(2k_2-k_1)/3}
	\right)
\end{equation}
are built on spin operators 1, 2, and 3 in the unit cell (see Fig.~\ref{lattfig}(a)), 
${\bf k} = k_1{\bf f}_1 + k_2{\bf f}_2$, and ${\bf f}_{1,2}$ are depicted in Fig.~\ref{lattfig}(b). 

\subsection{Harmonic approximation}

\begin{figure}
\includegraphics[scale=1.0]{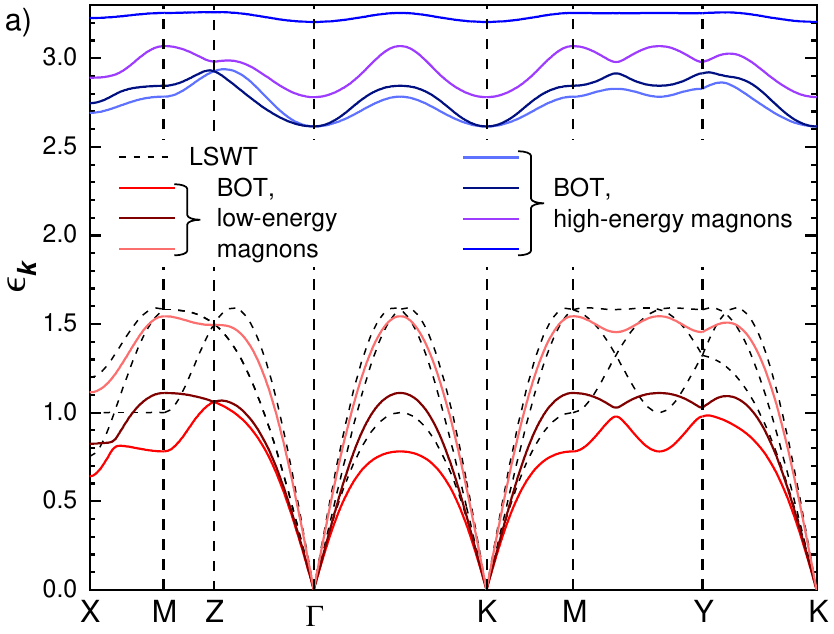}
\includegraphics[scale=1.2]{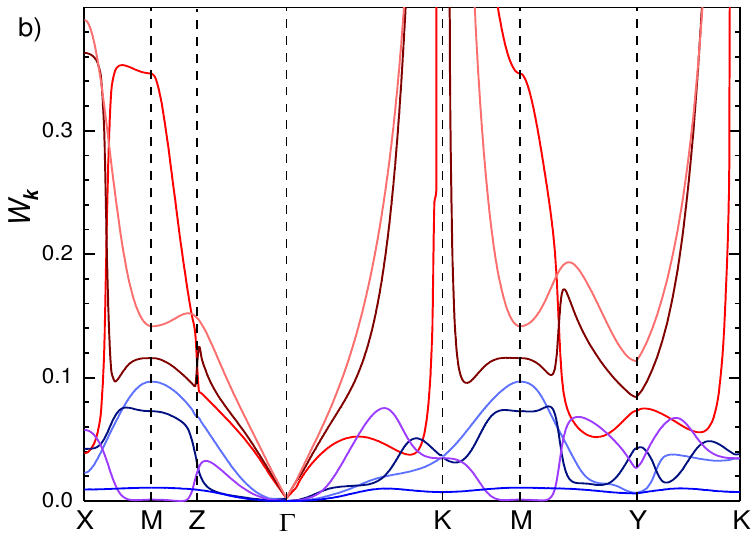}
\caption{
a) Spectra of elementary excitations corresponding to poles of dynamical spin susceptibility \eqref{chi} obtained in the linear spin-wave theory (LSWT) and within the harmonic approximation of the bond-operator technique (BOT). Three magnon branches in the LSWT correspond to the magnon spectrum $\epsilon_{\bf k}$, $\epsilon_{\bf k+k_0}$, and $\epsilon_{\bf k-k_0}$, where $\bf k_0$ is the antiferromagnetic vector. The path along the Brillouin zone is shown in Fig.~\ref{lattfig}(b).
b) Spectral weights $W_{\bf k}$ of the poles (i.e., coefficients before delta-functions in Eq.~\eqref{dsf}) found in the harmonic approximation of the BOT.
\label{spec0}}
\end{figure}

Spectra are shown in Fig.~\ref{spec0}(a) of seven branches of elementary excitations found in the harmonic approximation of the BOT. These excitations correspond to poles of spin correlator \eqref{chi} which is a linear combination of Green's functions of bosons in the harmonic approximation. Spectral weights of these poles (i.e., coefficients before delta-functions in Eq.~\eqref{dsf}) are presented in Fig.~\ref{spec0}(b). It is seen that contributions are significant of all quasiparticles to the spin correlator. That is why we call all of them "magnons" below. Fig.~\ref{spec0}(a) shows that one can distinguish three low-energy Goldstone excitations (low-energy magnons) and four high-energy branches (high-energy magnons). We examine in Appendix~\ref{polar} in some detail the polarization of these excitations. It is demonstrated there that the highest-energy low-energy magnon and the highest-energy high-energy magnon correspond to spins fluctuations transverse to staggered magnetization. The rest branches are of mixed nature contributing both to the transverse and to the longitudinal spin fluctuations (interestingly, the character of some of them changes upon passing along the BZ). This should be contrasted with the linear spin-wave theory (LSWT) in which all magnons are transverse quasiparticles. 

The magnon spectrum obtained in the LSWT is also presented in Fig.~\ref{spec0}(a). It can be found in two equivalent ways: (i) by introducing local rotating coordinate system at each lattice site and using the Holstein-Primakoff transformation with one type of Bose operators and momenta lying in the crystal (extended) BZ; \cite{chub_triang,zh_triang,zhito} (ii) by using the Holstein-Primakoff transformation for each spin in the magnetic unit cell (i.e., by introducing three types of Bose operators) and momenta lying in the magnetic BZ (see Fig.~\ref{lattfig}(b)) \cite{PhysRevB.40.2727}. Spectra of three magnon branches obtained using variant (ii) are equal to $\epsilon_{\bf k}$, $\epsilon_{{\bf k}+{\bf k}_0}$, and $\epsilon_{{\bf k}-{\bf k}_0}$, where $\epsilon_{\bf k}$ is the spectrum obtained in way (i) and ${\bf k}_0$ is an antiferromagnetic vector (${\bf k}_0$ equal to ${\bf f}_1$ and $-{\bf f}_1$ describes $120^\circ$ magnetic structures with different chiral orders, where ${\bf f}_1$ is shown in Fig.~\ref{lattfig}(b) \cite{TAF}). 

It is clear that three (Goldstone) magnon branches in the LSWT correspond to three (Goldstone) low-energy magnons in the BOT. It is seen from  Fig.~\ref{spec0}(a) that the spectrum found in the BOT is shifted down noticeably compared with the result of the LSWT. Besides, the amount of short-range quantum fluctuations taken into account in the harmonic approximation of the BOT lifts the classical spectrum degeneracy (i.e., the degeneracy of two out of three branches $\epsilon_{\bf k}$, $\epsilon_{{\bf k}+{\bf k}_0}$, and $\epsilon_{{\bf k}-{\bf k}_0}$) arising at blue dashed lines and at $\Gamma M$ lines in Fig.~\ref{lattfig}(b). From the point of view of the SWT, it may seem that the magnon spectrum degeneracy is robust against quantum fluctuations because first-order corrections in $1/S$ do preserve it. \cite{chub_triang,zh_triang,zhito} However, we demonstrate below that the lifting of the spectra degeneracy found in the BOT is confirmed quantitatively experimentally in $\rm Ba_3CoSb_2O_9$.

Due to the quantum nature of the considered system and cumbersomeness of the BOT, it is difficult to visualize somehow or to give a simple idea of the excited states arising in the BOT and compare them with their counterparts in the LSWT. Nevertheless, we try to do this in Appendix~\ref{excite} for the special point $Z$ in the BZ (see Fig.~\ref{lattfig}(b)). The point $Z$ is convenient for this purpose because there are no zero-point fluctuations within the LSWT and the zero-point fluctuations are very small in the harmonic approximation of the BOT.

\subsection{Spectrum in the first order in $1/n$}
\label{secfo}

\begin{figure}
\includegraphics[scale=1.0]{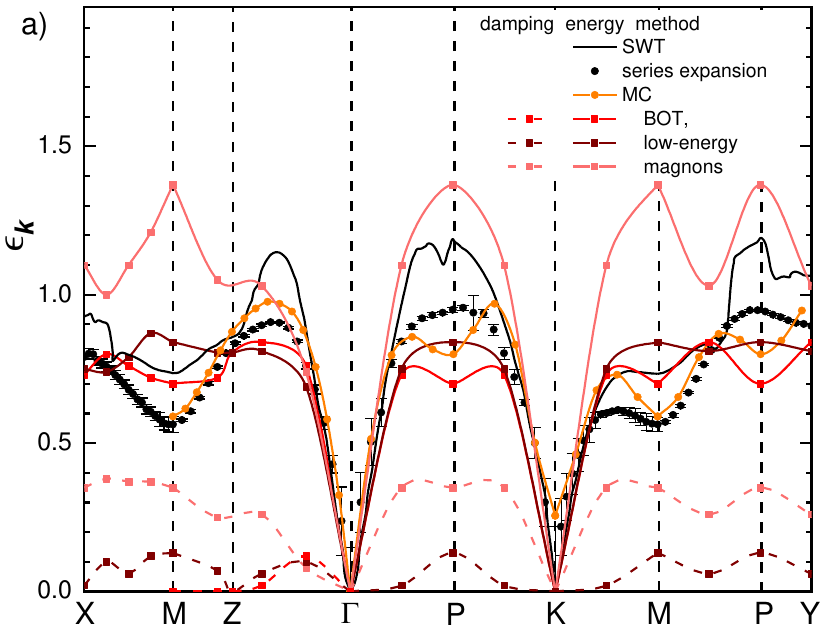}
\includegraphics[scale=1.0]{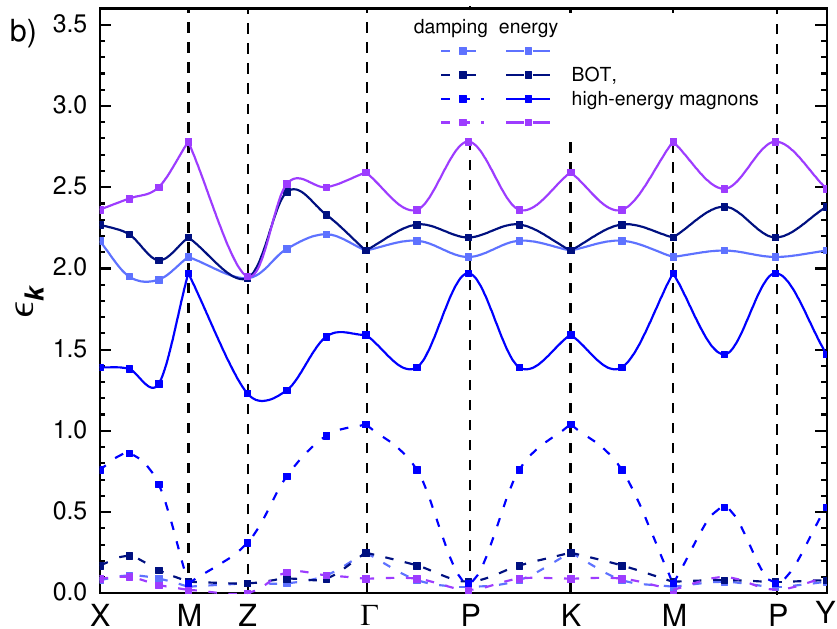}
\caption{
a) Spectra of the magnon branch obtained in the spin-wave theory (SWT) in the first order in $1/S$ (Ref.~\cite{chub_triang,chernprl}) and using the series-expansion technique (Ref.~\cite{series_tri}). Position of the low-energy magnon anomaly is shown which was observed using the dynamical variational Monte Carlo (MC) approach in Ref.~\cite{triang3}. Three low-energy magnon branches are presented obtained in the present study in the first order of the BOT (cf.\ Fig.~\ref{spec0}). These three branches in the BOT correspond to the magnon spectrum $\epsilon_{\bf k}$, $\epsilon_{\bf k+k_0}$, and $\epsilon_{\bf k-k_0}$ in the SWT and in the series-expansion technique ($\epsilon_{\bf k+k_0}$ and $\epsilon_{\bf k-k_0}$ are not available in the literature along the entire path and so they are not presented).
b) Spectra of four high-energy spin excitations obtained in the present study in the first order in $1/n$. Colors in both panels correspond to Fig.~\ref{spec0}(a).
\label{spec1}}
\end{figure}

We calculate now spectra of seven branches of excitations in the first order in $1/n$ in the standard way by expanding the denominator of spin correlator \eqref{chi} near bare poles $\epsilon_{i\bf k}^{(0)}$ ($i=1,\dots,7$) and taking self-energy parts at $\omega=\epsilon_{i\bf k}^{(0)}$. The results are presented in Fig.~\ref{spec1}. It is seen that all excitations except for the lowest-energy one acquire finite damping due to the decay into two other magnons. It is seen from Fig.~\ref{spec1}(a) that the spectrum of the well-defined lowest-energy magnon follows the position of the low-energy magnon anomaly observed using the dynamical variational Monte Carlo approach in Ref.~\cite{triang3}. In particular, the "roton" minima around $M$ and $P$ points are reproduced by the BOT. Our finding that the lowest-energy magnon is long-lived is in agreement with recent numerical results obtained by different methods \cite{Verresen2019,triang3}.

It is seen from Fig.~\ref{spec1}(b) that high-energy magnons acquire moderate damping except for the magnon shown in blue which is overdamped in the whole BZ apart from the vicinity of $M$ and $P$ points. We demonstrate in the next section that moderately damped high-energy magnons produce the high-energy anomaly in the DSF which was observed experimentally in $\rm Ba_3CoSb_2O_9$.

Notice also that the spectrum of short-wavelength quasiparticles found in the first order in $1/n$ self-consistently (i.e., by finding zeros of the denominator of spin correlators taking into account $\omega$-dependence of self-energy parts and not expanding the denominator near bare poles) differ from results of the present section (see below).

\subsection{Dynamical structure factor in the first order in $1/n$}

We turn to the calculation of the dynamical structure factor \eqref{dsf} at $M$ and $Y$ points by finding all self-energy parts in the first order in $1/n$ and taking into account their $\omega$-dependence. We stress that we do not expand in this section neither numerator nor denominator of spin correlator \eqref{chi}. By varying $n$ value, we have traced the evolution of the spin correlator poles from the limit $n\to\infty$ (harmonic approximation) to $n=1$. 

\begin{figure}
\includegraphics[scale=1.1]{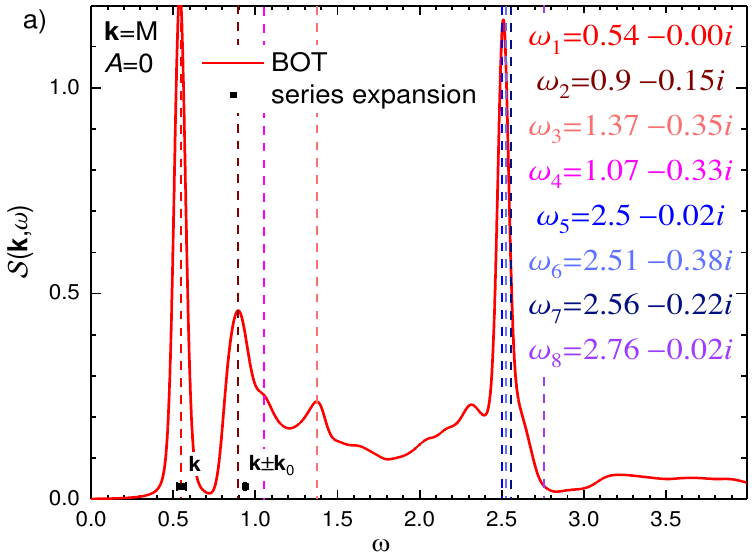}
\includegraphics[scale=1.1]{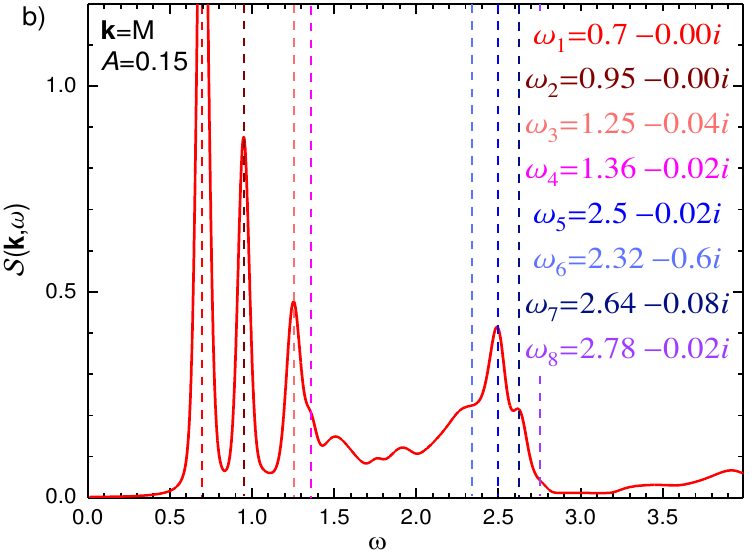}
\caption{
Dynamical structure factor (DSF) \eqref{dsf} at point $M$ of the BZ (see Fig.~\ref{lattfig}(b)) for spin-$\frac12$ antiferromagnet \eqref{ham} on the triangular lattice at a) $A=0$ and b) $A=0.15$. DSF obtained within the first order in $1/n$ has been convoluted with the energy resolution of $0.03J$. Magnon energies are also indicated in panel (a) which were obtained in Ref.~\cite{series_tri} using the series expansion technique. Anomalies in the DSF are produced by poles of spin correlator \eqref{chi} indicated in insets by colors corresponding to excitation branches shown in Fig.~\ref{spec0}(a). Real parts of these poles are marked by vertical dashed lines of respective colors. Imaginary parts of poles correspond to quasiparticles damping. Pole $\omega_4$ has no counterpart neither in the spin-wave theory nor in the harmonic approximation of the BOT.
\label{dsfm}}
\end{figure}

The result is shown in Fig.~\ref{dsfm}(a) for $M$ point at $n=1$. Poles of the correlator are also indicated in the inset of Fig.~\ref{dsfm}(a) by colors corresponding to Fig.~\ref{spec0}(a) (imaginary parts of poles give quasiparticles damping). Notice that these poles values are found in the self-consistent way so that the results differ from our findings from Sec.~\ref{secfo} (see Fig.~\ref{spec1}). The difference is small for low-energy magnons whereas it reaches 25\% for some high-energy magnons (cf.\ Figs.~\ref{spec1} and \ref{dsfm}(a)). It is seen from Fig.~\ref{dsfm}(a) that our self-consistent findings are in excellent agreement with previous numerical results obtained in Ref.~\cite{series_tri} using the series expansion. We point out that an incoherent background arises in the DSF at $\omega\agt0.7$ due to the two-magnon decay so that the lowest-energy magnon is well-defined and the rest magnons acquire finite damping and produce anomalies mounted on the incoherent background. Importantly, three high-energy magnons one of which has a very small damping give the high-energy anomaly at $\omega\approx2.5$ (at $\omega\approx2$ according to Fig.~\ref{spec1}(b)) which is observed in the experiment as we demonstrate below.

\begin{figure}
\includegraphics[scale=1.1]{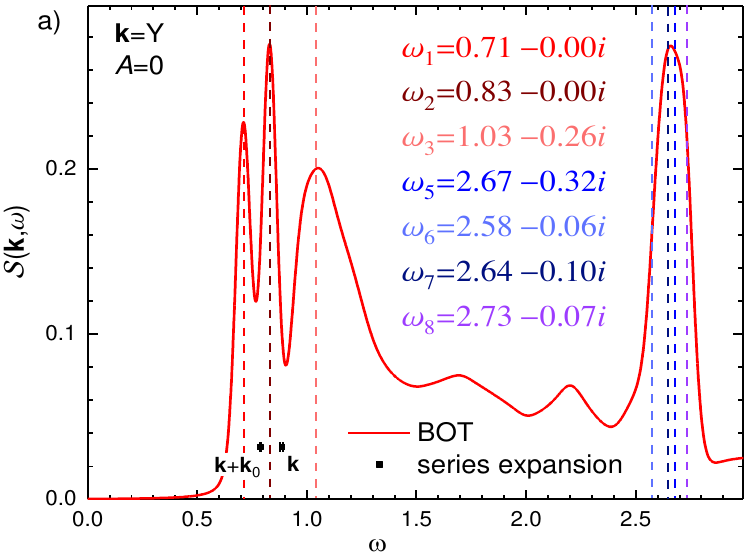}
\includegraphics[scale=1.1]{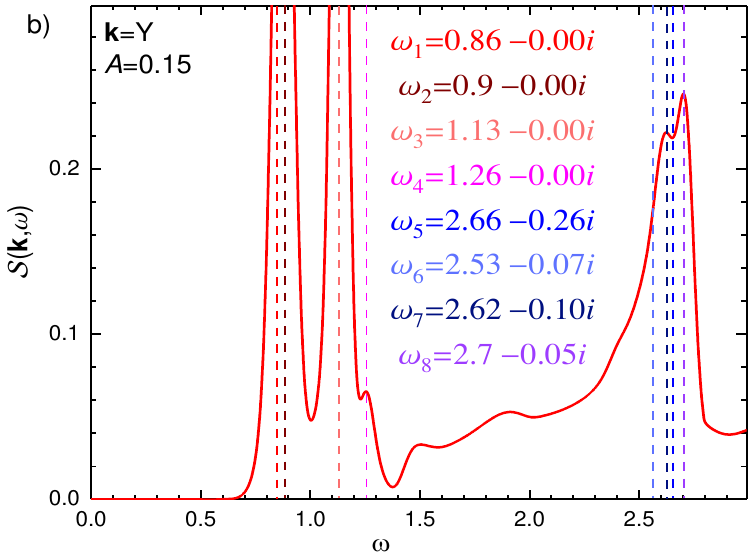}
\caption{
The same as Fig.~\ref{dsfm} but for point $Y$ of the BZ (see Fig.~\ref{lattfig}(b)). Small easy-plane anisotropy reduces the phase space for the magnon decay so that the broad anomaly around $\omega\approx1.1$ in panel (a) turns into two resolution-limited peaks in panel (b).
\label{dsfy}}
\end{figure}

DSF behaves similarly near $Y$ point as it is seen from Fig.~\ref{dsfy}(a). The difference with the $M$ point is that the incoherent background starts at $\omega\approx0.9$, all four high-energy magnons contribute to the high-energy anomaly at $\omega\approx2.7$, and pole $\omega_4$ acquires a very large damping thus producing no anomaly in the DSF.

Interestingly, pole $\omega_4$ corresponds to the novel quasiparticle which arises near $M$ at $n\approx2$ (despite its quite large damping at $A=0$, it produces a visible anomaly in the DSF near $M$ as is seen in Fig.~\ref{dsfm}(a)). It has no counterparts neither in the SWT nor in the harmonic approximation of the BOT. Notice that this new elementary excitation arises only around $M$ and $P$ points at $A=0$: the imaginary part of the new pole increases quickly upon going away from these points and the corresponding anomaly in the DSF merges into the incoherent continuum as it is illustrated by Fig.~\ref{dsfy}(a) for $Y$ point. However small easy-plane anisotropy decreases drastically the damping of four low-energy quasiparticles so that pole $\omega_4$ becomes well-defined at $A=0.15$ as it is seen in Figs.~\ref{dsfm}(b) and \ref{dsfy}(b). We propose below that this new quasiparticle was observed experimentally in $\rm Ba_3CoSb_2O_9$ near $M$. It is interesting to point out that we have found by the BOT the appearance of new poles after taking into account self-energy parts only in another non-collinear spin-$\frac12$ system, HAF in strong magnetic field, \cite{iboth} while there were no such phenomena in models with collinear magnetic orderings \cite{ibot,aktersky}.

\section{Easy-plane anisotropy and comparison with experiment}
\label{anisotropy}

The main effect of the easy-plane anisotropy $A$ in Eq.~\eqref{ham} is the reduction of the phase space for a magnon to decay into two other magnons (this process is described by the diagram shown in Fig.~\ref{diag}(c)) as it was obtained before in the spin-wave analysis \cite{zhito,zh_triang}. It is seen from Figs.~\ref{dsfm} and \ref{dsfy} that imaginary parts of four low-energy poles are substantially reduced at $A=0.15$ at $M$ and $Y$ points. In particular, the anisotropy makes well-defined the quasiparticle corresponding to pole $\omega_4$ at both points. The anisotropy produces also a gap $\Delta$ in spectra of two low-energy magnons at $\Gamma$ and $K$ points.

As it was established before, $\rm Ba_3CoSb_2O_9$ is a perfect realization of model \eqref{ham} with $J\approx1.7$~meV, $A\approx0.1$, and a small exchange coupling between spins from nearest triangular planes $J'\approx0.05J$. \cite{bacoprl,bacogap,bacoH} We neglect the inter-plane interaction for simplicity and find a good agreement with experimental observations at
\begin{equation}
\label{param}
J=1.77\,{\rm meV}, \quad A=0.15.
\end{equation}
In particular, we obtain for the gap value at parameters \eqref{param} 
\begin{equation}
\label{gap}
	\Delta/J = 0.57n-0.18 \mapsto 0.39 \mbox{ (at }n=1)
\end{equation}
that is in a very good agreement with the experimental finding \cite{bacogap,bacoprb} $\Delta\approx0.7\,{\rm meV}\approx0.4J$.

To describe available neutron data, one has to calculate the following dynamical structure factor: \cite{Lowesey}
\begin{equation}
\label{neutron}
{\cal S}_n({\bf k},\omega) = 
\frac1\pi  {\rm Im}
\sum_{\alpha,\beta} 
\left(
\delta_{\alpha\beta} - \widehat k_\alpha \widehat k_\beta
\right)
\chi_{\alpha\beta}({\bf k},\omega),
\end{equation}
where $\alpha,\beta=x,y,z$, $\widehat{\bf k}={\bf k}/k$, and $\chi_{\alpha\beta}({\bf k},\omega)$ are spin correlators \eqref{chi} built on spin operators $S^\alpha$ and $S^\beta$. One has to take into account also that there are many domains in real samples of $\rm Ba_3CoSb_2O_9$ with different directions of staggered magnetization \cite{bacoprb} so that Eq.~\eqref{neutron} should be averaged over all such domains. 

\begin{figure}
\includegraphics[scale=1.1]{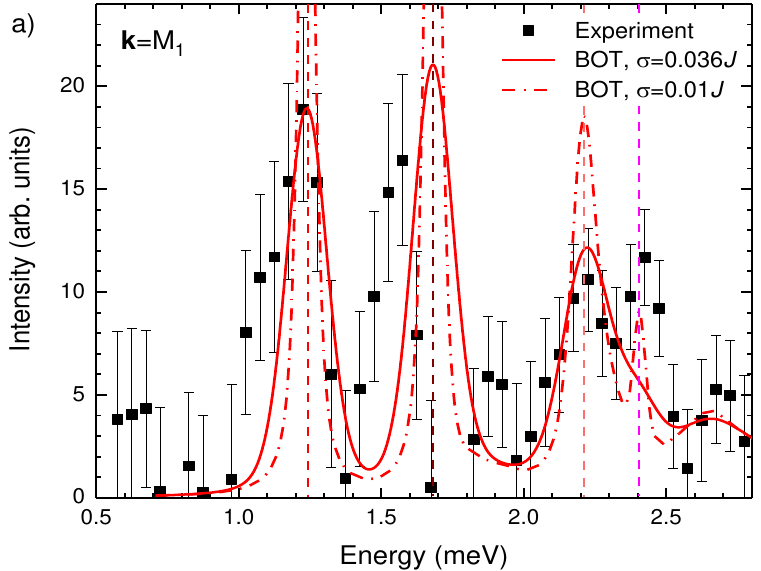}
\includegraphics[scale=1.1]{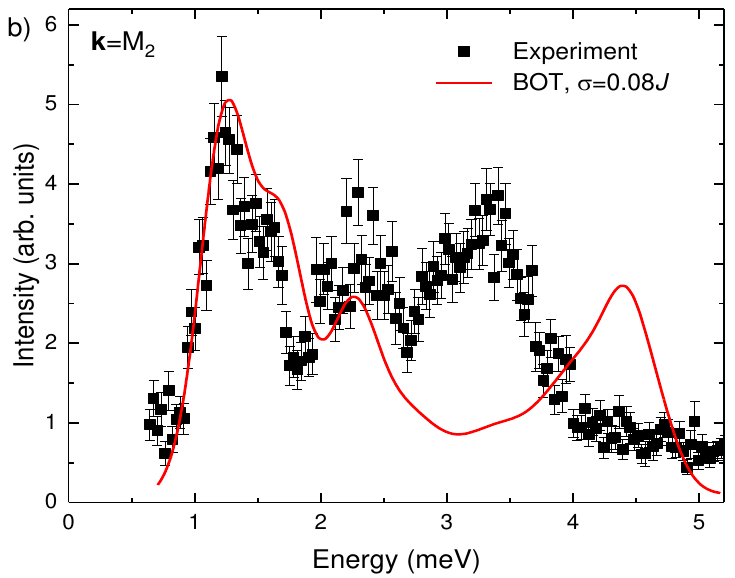}
\caption{Inelastic neutron scattering intensity with subtracted background obtained experimentally in $\rm Ba_3CoSb_2O_9$ at $M$ points ($M_1$ and $M_2$ correspond to ${\bf k}=(1/2,0,-1)$ and ${\bf k}=(1,1/2,-1)$, respectively). Red curves are theoretical results of the present study with parameters \eqref{param} in model \eqref{ham} convoluted with the energy resolution $\sigma$. (a) Experimental data from Ref.~\cite{bacoprl}, $\sigma=0.063\mbox{ meV}\approx0.036J$ (as in the experiment \cite{bacoprl}), and $\sigma=0.01J$. Vertical dashed lines indicate real parts of poles of spin correlator \eqref{neutron} as in Fig.~\ref{dsfm}(b). The pole shown in magenta has no counterpart neither in the spin-wave theory nor in the harmonic approximation of the BOT. (b) Experimental data from Ref.~\cite{triang1} and $\sigma=0.08J$ (as in the experiment \cite{triang1}). The discrepancy in the position of the high-energy anomaly between the theory and the experiment can be attributed to greater sensitivity of high-energy magnons to $1/n$ corrections as it is explained in the main text.
\label{experiment}}
\end{figure}

The result of our calculation of Eq.~\eqref{neutron} with parameters \eqref{param} is shown in Fig.~\ref{experiment}(a) at $M_1$ point (${\bf k}=(1/2,0,-1)$) together with experimental data from Ref.~\cite{bacoprl}. Four peaks are clearly seen in experimental data which are reproduced quite accurately by our results. The worse agreement is in the intensity of the peak at $\omega\approx2.4$~meV which corresponds to the novel quasiparticle described by pole $\omega_4$ in Fig.~\ref{dsfm}. The different ratio of the peaks spectral weights in Figs.~\ref{dsfm}(b) and \ref{experiment}(a) is accounted for by different weights of correlators $\chi_{xx}({\bf k},\omega)$, $\chi_{yy}({\bf k},\omega)$, and $\chi_{zz}({\bf k},\omega)$ in Eqs.~\eqref{dsf} and \eqref{neutron}. In particular, the low-energy peak at $M$ point is due to spin fluctuations in the direction transverse to the plain in which magnetic moments lie so that its spectral weight is zero in $\chi_{xx}({\bf k},\omega)$ and $\chi_{zz}({\bf k},\omega)$. This explains the increasing of the low-energy magnon energy as $A$ rises (see Figs.~\ref{dsfm}(a) and \ref{dsfm}(b)) and the diminishing of its spectral weight in Eq.~\eqref{neutron} upon increasing of the last component of $\bf k$.

Our results and data of another experiment \cite{triang1} performed on $\rm Ba_3CoSb_2O_9$ in a wider energy range (but with twice as bad energy resolution) than that in Ref.~\cite{bacoprl} are shown in Fig.~\ref{experiment}(b). The broad high-energy anomaly found experimentally around $\omega\approx3.4$~meV corresponds to the anomaly produced in our results by high-energy magnons. The discrepancy of 22\% in the position of this feature between the theory and the experiment can be attributed to greater sensitivity of high-energy poles of spin correlators to $1/n$ corrections: real parts of poles $\omega_{5,6,7,8}$ found self-consistently in Fig.~\ref{dsfm}(a) are about 25\% as large as corresponding values obtained using the denominator expansion in Fig.~\ref{spec1}(b). This signifies also that further $1/n$ corrections give a noticeable contribution to the renormalization of high-energy magnons spectra.

Notice that the SWT predicts only two magnon peaks at $M$ (due to the magnon spectra degeneracy discussed above) and a high-energy continuum of excitations \cite{chub_triang,zh_triang,zhito} whereas the BOT reproduces the number and positions of experimentally observed anomalies. We stress also that the good agreement with experiment confirms our finding that quantum fluctuations lift the degeneracy between two low-energy magnon branches predicted by the SWT along $\Gamma M$ lines and along blue dashed lines drawn in Fig.~\ref{lattfig}(b).

\section{Conclusion}
\label{conc}

To conclude, we use the three-spin variant of the BOT for discussion of spin dynamics in spin-$\frac12$ HAF on the triangular lattice. Our theory takes into account all excited states in the magnetic unit cell containing three spins and it respects the symmetry of the magnetic ordering (see Fig.~\ref{lattfig}). The ground-state energy and the value of the sublattices magnetization found in the first order in $1/n$ (see Eqs.~\eqref{statics}) are in good quantitative agreement with previous analytical and numerical findings.

We obtain seven branches of excitations in the BOT three of which are Goldstone quasiparticles ("low-energy magnons") known from the SWT and the rest four previously unknown branches ("high-energy magnons") originate from high-energy excitations of the unit cell. We find also the eighth (novel) quasiparticle which has no counterparts neither in the SWT nor in the harmonic approximation of the BOT and which has small enough damping around $M$ and $P$ points of the BZ (see Fig.~\ref{dsfm}). Similar to new elementary excitations obtained in Ref.~\cite{iboth} in HAF on the square lattice in strong magnetic field, the origin of the eighth quasiparticle is in strong quantum fluctuations in the system. We demonstrate that all observed quasiparticles produce visible anomalies in dynamical spin correlators. Spectra of low-energy magnons are in good agreement with previous numerical results. In particular, the BOT reproduces the "roton" minima in the spectrum of the well-defined low-energy magnon around $M$ and $P$ points. We show that in agreement with experiments in $\rm Ba_3CoSb_2O_9$ quantum fluctuations lift the degeneracy of two low-energy magnon branches predicted by the SWT along $\Gamma M$ lines and along blue dashed lines depicted in Fig.~\ref{lattfig}(b). High-energy magnons produce the broad high-energy anomaly in dynamical spin correlators as it is seen from Figs.~\ref{dsfm} and \ref{dsfy}.

In agreement with the conclusion made in the spin-wave analysis \cite{zhito,zh_triang}, we find that even small easy-plane anisotropy reduces considerably the phase space for magnon decay into two other magnons so that four low-energy elementary excitations obtained in the BOT have negligible damping at $A=0.15$ in Eq.~\eqref{ham}. We propose that four anomalies obtained in $\rm Ba_3CoSb_2O_9$ at $M_1$ point in Ref.~\cite{bacoprl} in the interval 0--3~meV stem from three low-energy magnons and the eighth quasiparticle (see Fig.~\ref{experiment}(a)). High-energy magnons found in the BOT contribute to the broad anomaly around 3.5~meV observed in $\rm Ba_3CoSb_2O_9$ in Ref.~\cite{triang1} (see Fig.~\ref{experiment}(b)). The discrepancy of 22\% in the position of this anomaly between the theory and the experiment can be attributed to the greater sensitivity of high-energy magnons to $1/n$ corrections and the necessity to go beyond the first order in $1/n$.

The easy-plane anisotropy produces the gap in spectra of two low-energy magnons at $K$ and $\Gamma$ points. The gap value given by Eq.~\eqref{gap} for model parameters \eqref{param} is in quantitative agreement with experimental findings in $\rm Ba_3CoSb_2O_9$.

Short-range spin correlations are taken into account more accurately in the BOT compared with standard approaches that results in the more precise description of the high-energy (short-wavelength) spin dynamics.

\begin{acknowledgments}

I am grateful to N.~Kurita and J.~Ma for exchange of data and useful discussions. 
This work is supported by the Russian Science Foundation (Grant No.\ 22-22-00028). 

\end{acknowledgments}

\appendix

\section{Basis for bond-operator theory}
\label{tecthe BOT}

\begin{figure}
\includegraphics[scale=0.6]{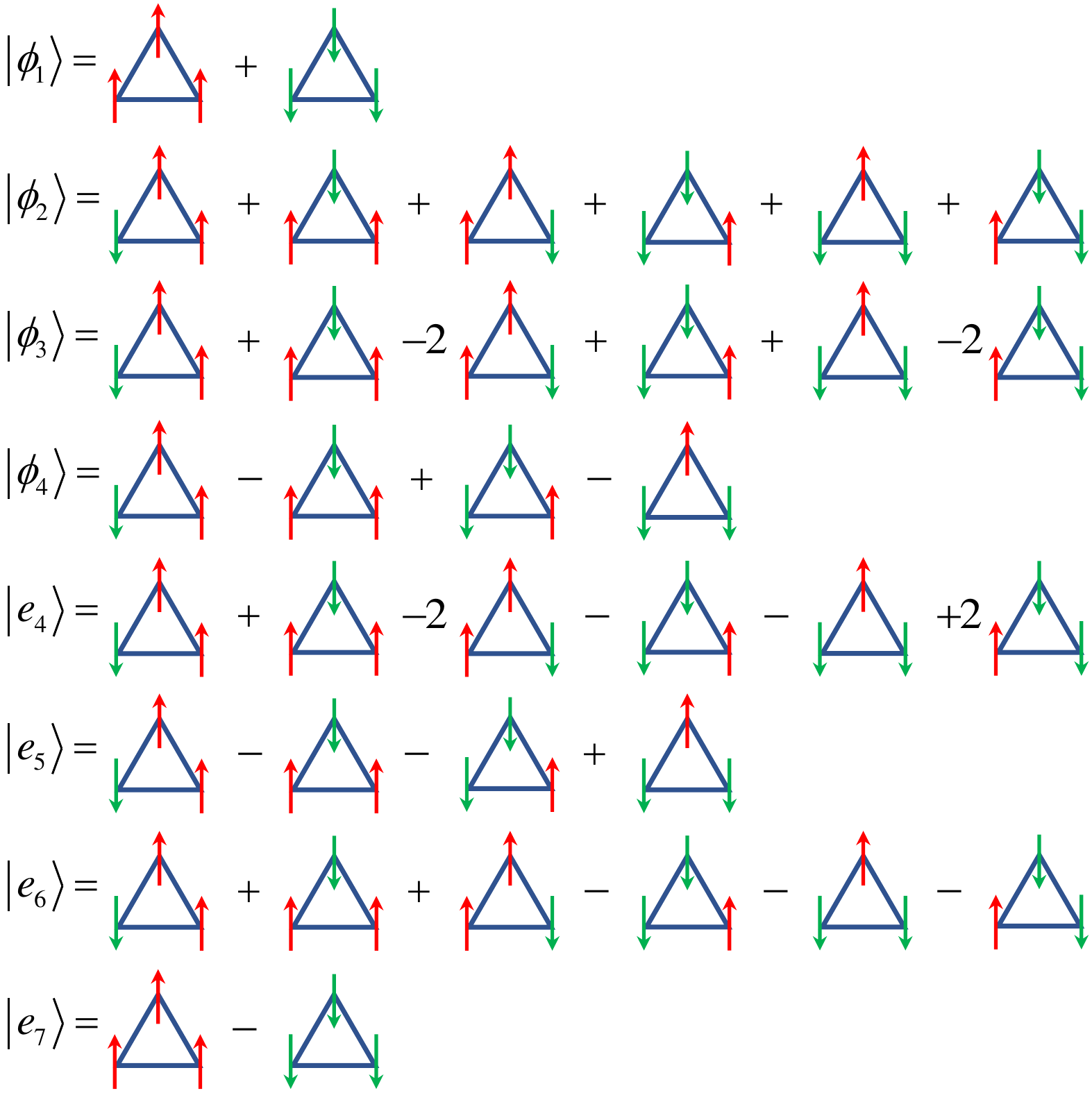}
\caption{Basis spin functions for the bond-operator technique. Normalization factors are omitted for clarity. 
\label{statesfig}}
\end{figure}

Basis functions for the proposed bond-operator theory (see Eq.~\eqref{bosons}) are shown in Fig.~\ref{statesfig}. All these states are simple linear combinations of eigenfunctions of the total spin $S$ and its $z$-projection $S^z$: $|\phi_1\rangle$ is the sum of the state with $(S=3/2,S^z=3/2)$ and the state with $(S=3/2,S^z=-3/2)$; $|\phi_{2,3,4}\rangle$ are sums of states with $(S=1/2,S^z=1/2)$ and states with $(S=1/2,S^z=-1/2)$; $|e_{4,5,6}\rangle$ are difference of states with $(S=1/2,S^z=1/2)$ and states with $(S=1/2,S^z=-1/2)$; $|e_7\rangle$ is the difference of the state with $(S=3/2,S^z=3/2)$ and the state with $(S=3/2,S^z=-3/2)$. Bearing in mind the common wisdom that the ground-state ordering is coplanar in the considered system and it is from the sector $S^z=0$, one can search the vacuum state $|0\rangle$ in the sector in which mean values are zero of operators $S_{1,2,3}^y$ and $S_1^z+S_2^z+S_3^z$, i.e., in the subspace formed by $|\phi_{1,2,3,4}\rangle$. Then, we confined ourselves to searching a coplanar magnetic ground-state ordering in the $xz$ plane because the plane in which spins lie do not effect the dynamics in the Heisenberg system. Then, it is convenient to represent $|0\rangle$ and $|e_{1,2,3}\rangle$ in Eq.~\eqref{bosons} as follows:
\begin{eqnarray}
\label{basfun}
|0\rangle &=&	
\cos\alpha_3 \left( |\phi_2\rangle \sin \alpha_1 + |\phi_1\rangle \cos\alpha_1 \right)
+
\sin\alpha_3 \left( |\phi_4\rangle \sin\alpha_2 + |\phi_3\rangle \cos\alpha_2 \right),
\nonumber\\
|e_1\rangle &=&	
\sin\alpha_3 \left( |\phi_2\rangle \sin\alpha_1 + |\phi_1\rangle \cos\alpha_1 \right)
-
\cos\alpha_3 \left( |\phi_4\rangle \sin\alpha_2 + |\phi_3\rangle \cos\alpha_2 \right),
\nonumber\\
|e_2\rangle &=&	
\cos\alpha_3 \left( |\phi_2\rangle \cos\alpha_1 - |\phi_1\rangle \sin\alpha_1 \right)
+
\sin\alpha_3 \left( |\phi_4\rangle \cos\alpha_2 - |\phi_3\rangle \sin\alpha_2 \right),
\\
|e_3\rangle &=&	
\sin\alpha_3 \left( |\phi_2\rangle \cos\alpha_1 - |\phi_1\rangle \sin\alpha_1 \right)
-
\cos\alpha_3 \left( |\phi_4\rangle \cos\alpha_2 - |\phi_3\rangle \sin\alpha_2 \right),
\nonumber
\end{eqnarray}
where real parameters $\alpha_{1,2,3}$ should be found as a result of minimization of the system ground-state energy (i.e., the term without Bose operators in the Bose-analog of spin Hamiltonian \eqref{ham}) or, equivalently, from the requirement that the term ${\cal H}_1$ in the Hamiltonian linear in Bose operators should vanish. We find in this way
$\alpha _1 = 1.0472$, 
$\alpha _2 = 0.2618$, 
and
$\alpha _3 = 1.8635$. 
There are also $1/n$ corrections to these quantities coming from the contribution to ${\cal H}_1$ from terms in the Hamiltonian containing products of three Bose operators after making all possible couplings of two Bose operators. As a result, we find
$\alpha _1 = 1.0472+0/n$, 
$\alpha _2 = 0.2618+0/n$, 
and
$\alpha _3 = 1.8635-0.1247/n$. 
Because all terms in the Hamiltonian depend on $\alpha_{1,2,3}$, $1/n$ corrections to these parameters contribute to the renormalization of observables in the first order in $1/n$ and we have taken them into account in all our calculations. Notice that mean spin components calculated both in the zeroth and in the first orders in $1/n$ show the $120^\circ$ magnetic ordering in the ground state in the $xz$ plane (it is the minimization of the system ground-state energy with respect to $\alpha_{1,2,3}$ that fixes the angle of $120^\circ$ between mean spin components in the zeroth order in $1/n$). 

It is just for the sake of reduction of the number of parameters in the theory that we choose the subspace for $|0\rangle$ in which mean values are zero of operators $S_{1,2,3}^y$ and $S_1^z+S_2^z+S_3^z$ (i.e., the subspace formed by $|\phi_{1,2,3,4}\rangle$). The $120^\circ$ magnetic structure which we obtain is in agreement with all results found before by other methods that corroborates our simplified consideration. The stability of the spectra of elementary excitations which we observe also indicates that the $120^\circ$ magnetic structure we find provides a minimum of the system energy. One has to introduce a linear combination of all basis functions for $|0\rangle$ (with complex coefficients) and minimize the system ground state energy with respect to all (complex) coefficients in order to discuss the ground-state ordering rigorously. This is a tedious procedure which has to be done in an unknown system whereas the consideration can be simplified by using previous results in the case of well-studied models. We choose the simplest way in the present study.

It should be stressed also that the BOT built on a basis containing linear combinations of all states $|\phi_{1,2,3,4}\rangle$ and $|e_{4,5,6,7}\rangle$ with 7 parameters $\alpha$ does give the same results for observables. One can consider also linear combinations of states $|e_{1,2,3,4,5,6,7}\rangle$ for excited states. However, Bose operators arisen in this case would be related with Bose operators \eqref{bosons} via a unitary transformation that guarantees the same results for observable quantities.

\section{Polarization of spin excitations}
\label{polar}

In this appendix, we explore the nature of seven spin excitations arising in the harmonic approximation of the BOT (see Fig.~\ref{spec0} for their spectra). It is well known that conventional magnons in the spin-wave theory are collective excitations related with fluctuations transverse to staggered magnetizations. In the longitudinal channel, there can arise other excitations one of which is the amplitude (Higgs) mode. As the longitudinal and the transverse channels are mixed in non-collinear magnets, one expects that all excitations would be of mixed nature. However, it is interesting to consider this point in some detail and discuss the spin susceptibility (cf.\ Eqs.~\eqref{chi} and \eqref{sk})
\begin{eqnarray}
\label{chit}
&&\chi^T({\bf k},\omega) =
i\int_0^\infty dt 
e^{i\omega t}	
\left\langle \left[ B_{\bf k}(t), C_{-\bf k}(0) \right] \right\rangle,\\
&&B_{\bf k} =
\frac{1}{\sqrt3} \left(
\left( \frac{\sqrt3}{2}S^x_{1\bf k} + \frac{1}{2}S^z_{1\bf k} + iS^y_{1\bf k} \right) 
- 
\left( S^z_{2\bf k} - iS^y_{2\bf k} \right) e^{-i(k_1+k_2)/3} 
+ 
\left( -\frac{\sqrt3}{2}S^x_{3\bf k} + \frac{1}{2}S^z_{3\bf k} + iS^y_{3\bf k} \right)  e^{-i(2k_2-k_1)/3}
\right),\nonumber\\
&&C_{\bf k} =
\frac{1}{\sqrt3} \left(
\left( \frac{\sqrt3}{2}S^x_{1\bf k} + \frac{1}{2}S^z_{1\bf k} - iS^y_{1\bf k} \right) 
- 
\left( S^z_{2\bf k} + iS^y_{2\bf k} \right) e^{-i(k_1+k_2)/3} 
+ 
\left( -\frac{\sqrt3}{2}S^x_{3\bf k} + \frac{1}{2}S^z_{3\bf k} - iS^y_{3\bf k} \right)  e^{-i(2k_2-k_1)/3}
\right),\nonumber
\end{eqnarray}
where $B_{\bf k}$ and $C_{\bf k}$ are built, respectively, on operators $S^+_{1,2,3}$ and $S^-_{1,2,3}$ in the local coordinate frames in which $z$ axes are directed along the local mean magnetizations. Here, we take into account also that for parameters $\alpha_{1,2,3}$ presented in Appendix~\ref{tecthe BOT} the mean magnetic moments are directed along $(-1/2,\sqrt3/2)$, $(1,0)$, and $(-1/2,-\sqrt3/2)$ at sites 1, 2, and 3 in the unit cell (see Fig.~\ref{lattfig}(a)), respectively. 

\begin{figure}
\includegraphics[scale=1.1]{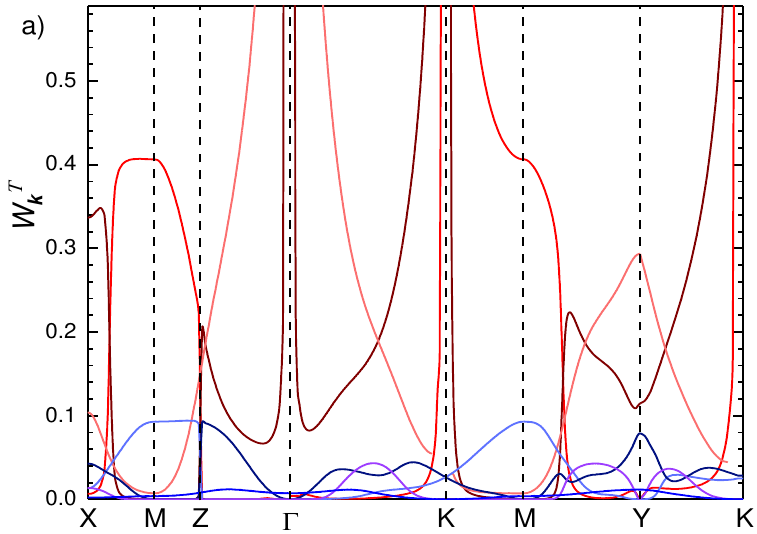}
\includegraphics[scale=1.1]{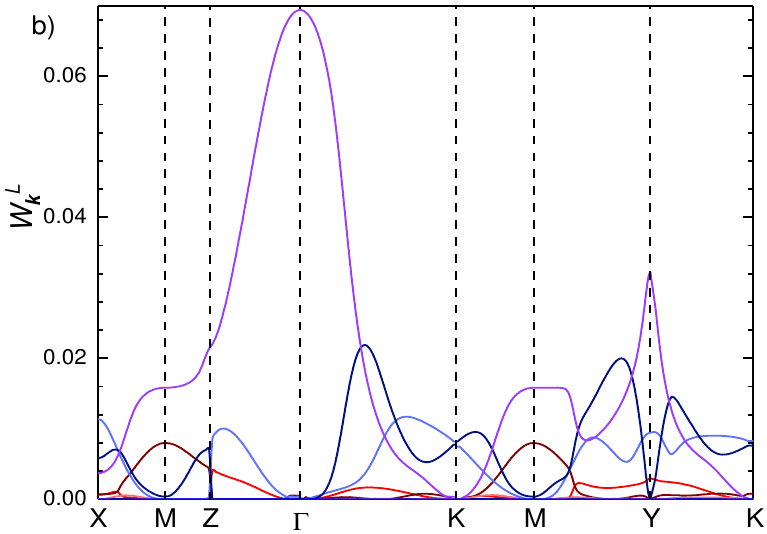}
\caption{Spectral weights $W_{\bf k}^{T,L}$ of peaks in the "transverse" and the "longitudinal" dynamical structure factors 
$
\frac1\pi {\rm Im}
\chi^{T,L}({\bf k},\omega)
$ 
built on correlators \eqref{chit} and \eqref{chil} which are calculated in the harmonic approximation of the BOT. 
\label{weightspol}}
\end{figure}

Similarly, we introduce the longitudinal spin correlator which is built on operators $S^z_{1,2,3}$ in the local coordinate frames
\begin{eqnarray}
\label{chil}
&&\chi^L({\bf k},\omega) =
i\int_0^\infty dt 
e^{i\omega t}	
\left\langle \left[ B_{\bf k}(t), C_{-\bf k}(0) \right] \right\rangle,\\
&&B_{\bf k} = C_{\bf k} =
\frac{1}{\sqrt3} \left(
\left( -\frac{1}{2}S^x_{1\bf k} + \frac{\sqrt3}{2}S^z_{1\bf k} \right) 
+ 
S^x_{2\bf k} e^{-i(k_1+k_2)/3} 
- 
\left( \frac{1}{2}S^x_{3\bf k} + \frac{\sqrt3}{2}S^z_{3\bf k} \right)  e^{-i(2k_2-k_1)/3}
\right).\nonumber
\end{eqnarray}
We plot in Fig.~\ref{weightspol} spectral weights of peaks in dynamical structure factors 
$
\frac1\pi {\rm Im}
\chi^{T,L}({\bf k},\omega)
$ 
built on correlators \eqref{chit} and \eqref{chil} which are calculated in the harmonic approximation of the BOT. It is seen that the highest-energy low-energy magnon and the highest-energy high-energy magnon are purely transverse. The rest branches are of mixed type and the character of some of them changes upon passing along the BZ.

\section{Spin excitations in the BOT and in the spin-wave theory}
\label{excite}

We try to give a simple idea in this appendix of excited states arising in the BOT and we try to compare them with their counterpart in the linear spin-wave theory (LSWT). This is particularly easy to do at the special point $Z$ of the BZ (see Fig.~\ref{lattfig}(b)) at which there are no zero-point fluctuations within the LSWT and they are very small in the harmonic approximation of the BOT. For the sake of comparison, it is convenient to develop the standard LSWT with three types of bosons (i.e., by performing the Holstein-Primakoff transformation at each site in the magnetic unit cell in the local coordinate frame).  

\begin{figure}
\includegraphics[scale=0.6]{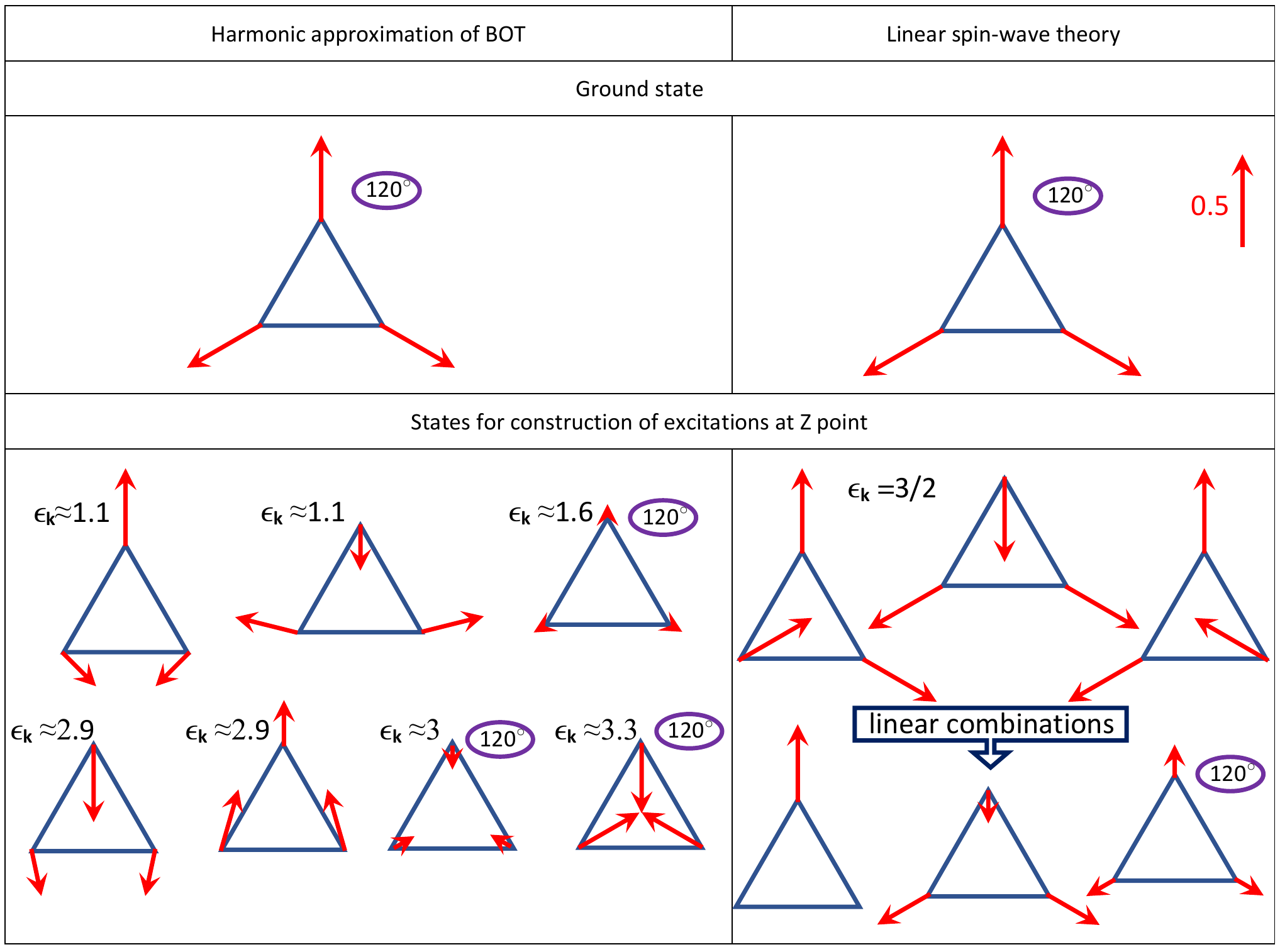}
\caption{Mean values $\langle {\bf S}_{1,2,3}\rangle$ of three spins in the magnetic unit cell in the ground state and in quantum states which can be used for the construction of excitations at $Z$ point in the BZ (see Fig.~\ref{lattfig}(b)) by operator $\frac{1}{\sqrt N}\sum_j e^{i{\bf k}{\bf R}_j} b_j^\dagger$, where $b_j^\dagger$ creates the corresponding state at $j$-th unit cell. There are no zero-point fluctuations at $Z$ point within the linear spin-wave theory; three excited states are built on simple spin flips in magnetic unit cells which have energy $\epsilon_{\bf k}=3/2$. Zero-point fluctuations are small at $Z$ point within the harmonic approximation of the BOT and we neglect them here. Components are zero of $\langle {\bf S}_{1,2,3}\rangle$ perpendicular to the plane of the figure.
\label{excitefig}}
\end{figure}

Due to the absence of zero-point fluctuations, the contribution to the Hamiltonian from $Z$ point has a simple form in the LSWT
$
\epsilon_{\bf k}(b_{1\bf k}^\dagger b_{1\bf k} + b_{2\bf k}^\dagger b_{2\bf k} + b_{3\bf k}^\dagger b_{3\bf k})
$, 
where $\epsilon_{\bf k}=3/2$, and the excited states with the corresponding $\bf k$ are created from the classical ground state having $120^\circ$ magnetic structure by operators $\frac{1}{\sqrt N}\sum_j e^{i{\bf k}{\bf R}_j} b_{qj}^\dagger$, where $N$ is the number of unit cells in the lattice, $q=1,2,3$, and $b_{qj}^\dagger$ creates a simple spin flip at $q$-th site in the $j$-th unit cell (see Fig.~\ref{excitefig}).

Within the harmonic approximation of the BOT, by discarding terms in the Hamiltonian describing zero-point fluctuations (i.e., terms containing products of two operators of creation or two operators of annihilation), one can bring the contribution to the Hamiltonian from $Z$ point to the simple form
$
\sum_{q=1}^7\epsilon_{q\bf k}b_{q\bf k}^\dagger b_{q\bf k} 
$,
where $(b_1,b_2,...,b_7) = U (a_1,a_2,...,a_7)$, $a_i$ are introduced in Eq.~\eqref{bosons}, and $U$ is a unitary matrix. States created by $b_{qj}^\dagger$ are related with basis functions shown in Fig.~\ref{statesfig} by the unitary transformation $U$. Fig.~\ref{excitefig} demonstrates mean spin values in these states, where corresponding $\epsilon_{q\bf k}$ are also presented. The latter values are very close to the bare spectrum at $Z$ point shown in Fig.~\ref{spec0} that indicates the minor role of zero-point fluctuations at this momentum.

\bibliography{tribib}

\end{document}